\documentclass[11pt,onecolumn]{article}
\usepackage{amssymb}
\usepackage{amsmath}
\usepackage[dvipdfmx]{graphicx}
\usepackage{cite}
\begin{document}





\begin{center}

{\large 
{\bf Bottom-tau unification by neutrinos in a nonsupersymmetric SU(5) model} 
}

\vskip 0.3cm

Takanao Tsuyuki\\

\vskip 0.3cm

{ \it Graduate School of Science and Technology, Niigata University, Niigata 950-2181, Japan}

\begin{abstract} 
We show that Yukawa couplings of bottom quarks and tau leptons can be unified in a non-supersymmetric SU(5) model. We introduce an arbitrary number of right-handed neutrinos. Their masses and Yukawa couplings that satisfy the unification condition by renormalization group evolution are shown. In the case that the grand unification scale is $10^{15.5}$GeV and three right-handed neutrinos have the same mass, the upper bound on their mass is $\sim 10^{14.1}$GeV.
\end{abstract}

\end{center}

{\it Introduction.} Neutrino oscillations, which mean that neutrinos have masses, are evidence for physics beyond the Standard Model (SM). We know that at least two flavors of neutrinos have $O$(0.01-0.1)eV masses. Their Majorana masses can be written as $v^2/M$ ($v\simeq174$GeV). We need $M$ to be in the desert between the electroweak and Planck scales, $O(10^{15})$ GeV (seesaw scale). The neutrino masses can be induced by singlet, right-handed neutrinos (type-I \cite{Minkowski:1977sc,GellMann:1980vs,Yanagida:1979as}), an SU(2) triplet scalar (type-II \cite{Barbieri:1979ag,Cheng:1980qt,Schechter:1980gr }) or  SU(2) triplet fermions (type-III seesaw \cite{Foot:1988aq}). 

Near the seesaw scale, there may be another interesting phenomenon: unification of elementary forces\cite{Pati:1974yy,Georgi:1974sy}. If the SM gauge group $G_{\rm SM}\equiv \rm SU(3)\times SU(2)\times U(1)$ is embedded into one non-Abelian group, we can explain the charge quantization of quarks and leptons. The simplest grand unification theory (GUT) is the model based on the SU(5) gauge group\cite{Georgi:1974sy}, which is broken down to $G_{\rm SM}$ in one step. By solving renormalization group equations (RGEs), the three gauge couplings indeed evolve toward unification, but do not exactly meet at one scale. If some fields charged under $G_{\rm SM}$ exist between the electroweak scale and the GUT scale ($M_G$), unification is still possible. The cases of a scalar representation $15_H$ \cite{Dorsner:2005fq,Dorsner:2005ii} and a fermion representation $24_F$\cite{Krasnikov:1993sc,Bajc:2006ia,Bajc:2007zf,Ibe:2009gt} are especially interesting since they can also explain the neutrino masses by type-II  or type-I+III  seesaw mechanisms and even the baryon asymmetry of the Universe. 

In the SU(5) model, not only gauge couplings, but also eigenvalues of Yukawa coupling matrices of down-type quarks and charged leptons are unified at $M_G$. This unification condition cannot be satisfied if we consider known fields only \cite{Arason:1991ic}. In previous studies, a dimension five operator is assumed for this problem \cite{Ellis:1979fg,Dorsner:2006hw,Bajc:2006ia,Bajc:2007zf}. The contribution of the operator is $\lesssim vM_G/\Lambda$. $M_G$ is bounded from below by the nucleon decay search \cite{Nishino:2012ipa}, naively $M_G\gtrsim 10^{15.4}$GeV\cite{Ibe:2009gt}. The GUT scale near this bound is phenomenologically interesting and natural, because it may be tested by the nucleon decay search\cite{Abe:2011ts,Shiozawa:2013taa} and is close to the seesaw scale. If the GUT scale is $10^{15.5-16.0}$GeV, and $\Lambda\sim10^{19}$GeV, the higher-dimensional term gives $\lesssim 0.1\rm GeV$. This is suitable for adjusting the first- and the second-generation Yukawa coupling unifications (and may explain why the $u$ quark is lighter than the $d$ quark), but is too small for that of the third generation. We need some mechanism for $y_b=y_\tau$, i.e., $b-\tau$ unification. 

The main purpose of this letter is to show that $b-\tau$ unification is possible if we introduce right-handed neutrinos to a non-supersymmetric SU(5) model. If there are right-handed neutrinos, the Yukawa couplings of neutrinos $y_\nu$ changes the RGEs of Yukawa couplings of quarks and charged leptons \cite{Grzadkowski:1987tf,Antusch:2005gp}. 
$b-\tau$ unification is not possible in the of type-II and type-III seesaw cases because an SU(2) triplet scalar and triplet fermions contribute to the running of $y_\tau$ in the positive direction \cite{Dorsner:2006hw, Chakrabortty:2008zh}. We consider 1-loop beta functions throughout this letter mainly for a large experimental error of $m_b$. The gauge couplings are assumed to be unified by adjoint fermions and a scalar, but the details of gauge coupling unification do not essentially change our analysis of $b-\tau$ unification.

 \newpage

\begin{figure}[tb]
\begin{center}
 \includegraphics[width=13cm]{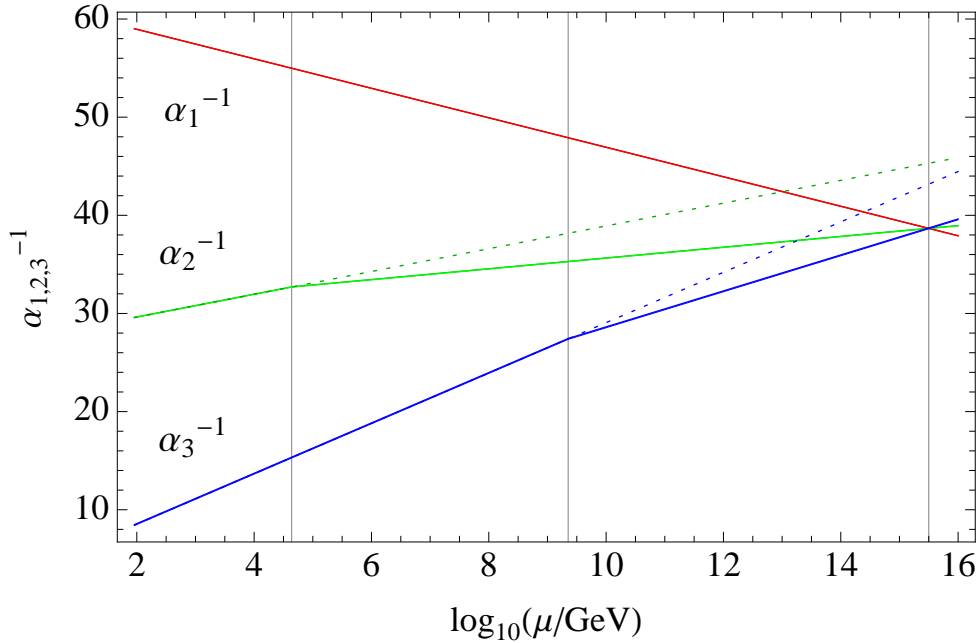}
\caption{Gauge coupling running considering one standard deviation. The GUT scale is taken to be $10^{15.5}$GeV. Dotted lines show the SM case ($\alpha_1$ does not change).}
\label{fgauge}
\end{center}
\end{figure}

{\it Gauge and Yukawa coupling unifications.} For gauge coupling unification, we assume that three multiplets $T_F,\;T_H\equiv(1,3,0),\;O_F\equiv(8,1,0)$ (indicating representations under $G_{\rm SM}$ with $F (H)$ meaning fermionic (Higgs) field) are much lighter than $M_G$. These fields can be embedded into adjoint representations of SU(5),
\footnote{We need at least two generations of the adjoint fermion $24_F$. If $T_F$ and $O_F$ are components of the same generation of $24_F$, the other components such as $(3,2,-5/6)$ acquire masses $\lesssim M_G^2/\Lambda$\cite{Bajc:2006ia,Bajc:2007zf,DiLuzio:2013dda}. These additional light fields change the RGEs and make gauge unification at $M_G\gtrsim 10^{15.5}$GeV impossible at the 1-loop level if $\Lambda$ is the Planck scale.}
$24_{F,H}=(1,1,0)+(1,3,0)+(8,1,0)+(3,2,-5/6)+(\bar{3},2,5/6)$.
 The singlet scalar in $24_{H}$ breaks SU(5) down to $G_{\rm SM}$. By the condition of gauge unification at $M_G$, we obtain
\begin{align}
(m_{T_F}^4 m_{T_H})^{1/5}&=M_Z\exp\left[\frac{2\pi}{\Delta b_2}(\alpha_{1,Z}^{-1}-\alpha_{2,Z}^{-1})+\frac{b_2'-b_1}{\Delta b_2}t_G\right]=4.37\times 10^4 {\rm GeV}\left(\frac{10^{15.5}\rm GeV}{M_G}\right)^\frac{84}{25},\\
m_{O_F}&=M_Z\exp\left[\frac{2\pi}{\Delta b_3}(\alpha_{1,Z}^{-1}-\alpha_{3,Z}^{-1})+\frac{b_3'-b_1}{\Delta b_3}t_G\right]=2.25\times 10^9 {\rm GeV}\left(\frac{10^{15.5}\rm GeV}{M_G}\right)^\frac{91}{20}.
\end{align}
Here we have used the experimental values  $\alpha=(127.940\pm0.014)^{-1}$, $\alpha_3=0.1185\pm0.0006$, 
$\sin^2\theta_W=0.23126\pm0.00005$ at $\mu=M_Z=91.1876$GeV \cite{Agashe:2014kda} ($\mu$ denotes the scale of renormalization). We have defined $(b_1,b_2',b_3')\equiv(41/10,-3/2,-5),\;(\Delta b_2, \Delta b_3)\equiv(5/3,2),\;t_G\equiv\ln(M_G/M_Z)$. The running of gauge couplings in the $M_G=10^{15.5}$GeV case is shown in Fig. \ref{fgauge}. The light triplets can be searched for in collider experiments \cite{Bajc:2006ia,Bajc:2007zf,CMS:2012ra,Aguilar-Saavedra:2013twa}. By a condition $m_{T_F}, m_{T_H}>M_Z$, we get $M_G<10^{16.3}$GeV. Note that this bound may be changed by higher loop corrections \cite{DiLuzio:2013dda}.

Next, we consider the running of Yukawa couplings. We assume that $N_g$ singlet ((1,1,0)) fermions are coupled to left-handed neutrinos by Yukawa couplings $Y_{\nu Ii}\;(I=1,\dots,N_g,i=1,2,3)$. We call these singlets right-handed neutrinos. They can be embedded into $24_F$ or other multiplets. For simplicity, we assume that $Y_{\nu I3}$ are the dominant components and their absolute values are the same, $|Y_{\nu I3}|=y_\nu$. We also consider the case in which all the masses of the right-handed neutrinos are the same $M_N$. The RGEs of the relevant Yukawa couplings at $\mu>M_N$  are \cite{Grzadkowski:1987tf,Antusch:2005gp}, 
\begin{align} 
16\pi^2 \frac{dy_t}{dt}&=y_t\left(\frac{3}{2}y_t^2-\frac{3}{2}y_b^2+T-\frac{17}{20}g_1^2-\frac{9}{4}g_2^2-8g_3^2\right), \label{eytop}\\
16\pi^2 \frac{dy_b}{dt}&=y_b\left(\frac{3}{2}y_b^2-\frac{3}{2}y_t^2+T-\frac{1}{4}g_1^2-\frac{9}{4}g_2^2-8g_3^2\right), \label{eyb}\\
16\pi^2 \frac{dy_\tau}{dt}&=y_\tau\left(\frac{3}{2}y_\tau^2-\frac{3}{2}N_gy_\nu^2+T-\frac{9}{4}g_1^2-\frac{9}{4}g_2^2\right), \label{eytau}\\
16\pi^2 \frac{dy_\nu}{dt}&=y_\nu\left(\frac{3}{2}N_gy_\nu^2-\frac{3}{2}y_\tau^2+T-\frac{9}{20}g_1^2-\frac{9}{4}g_2^2\right), \label{eyn}\\
T&\equiv 3y_t^2+3y_b^2+y_\tau^2+N_gy_\nu^2,
\end{align}
where we define $t\equiv \ln(\mu/M_Z)$. We also use $t_N\equiv\ln(M_N/M_Z)$ below. We have neglected Yukawa couplings of the first and second generations of quarks and charged leptons and the light $T_F$ since they are small ($y_{T_F}\lesssim \sqrt{m_{T_F}m_\nu}/v\sim10^{-5}$). $y_\nu$ contributes to $y_\tau$ negatively, so we can expect that large $y_\nu$ unifies $y_b$ and $y_\tau$. For initial conditions, we use $y_b(0)v=2.86^{+0.16}_{-0.06}$GeV, $y_\tau(0) v =1.74617$GeV, $y_t(0) v =172.1\pm 1.2$GeV \cite{Xing:2011aa}, where $v\equiv2^{-3/4}G_F^{-1/2},\;G_F=1.1663787 \rm GeV^{-2}$\cite{Agashe:2014kda}.

\begin{figure}[htb]
\begin{center}
\includegraphics[width=13cm]{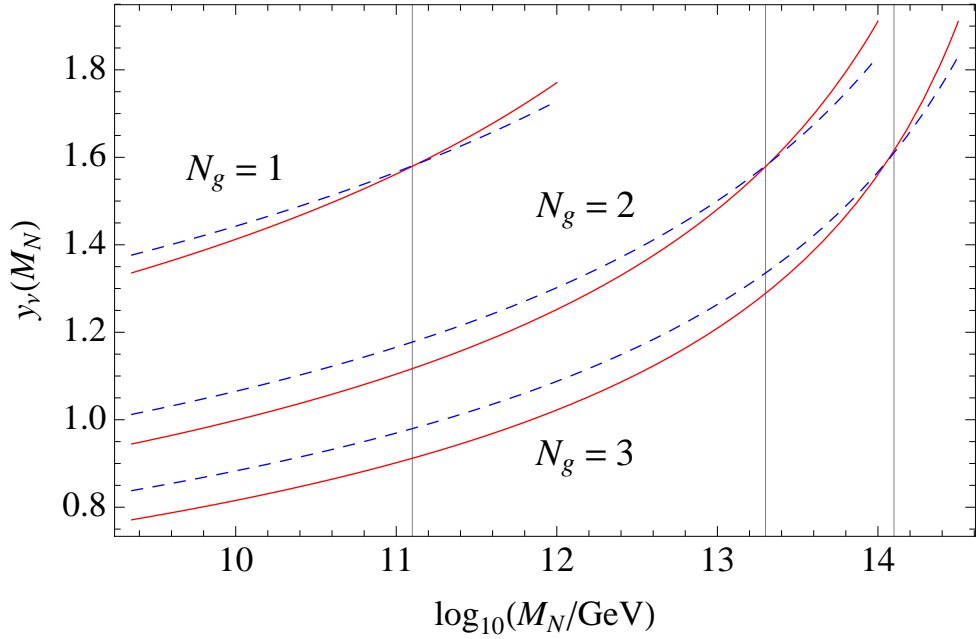}
\caption{Solid curves show the initial conditions of neutrino Yukawa couplings at $\mu=M_N$ that realize $b-\tau$ unification. Dashed curves show upper bounds from perturbativity ($M_G=10^{15.5}$GeV). Vertical lines indicate crossing points.}
\label{fynmn}
\end{center}
\end{figure}

\begin{figure}[htb]
\begin{center}
\includegraphics[width=13cm]{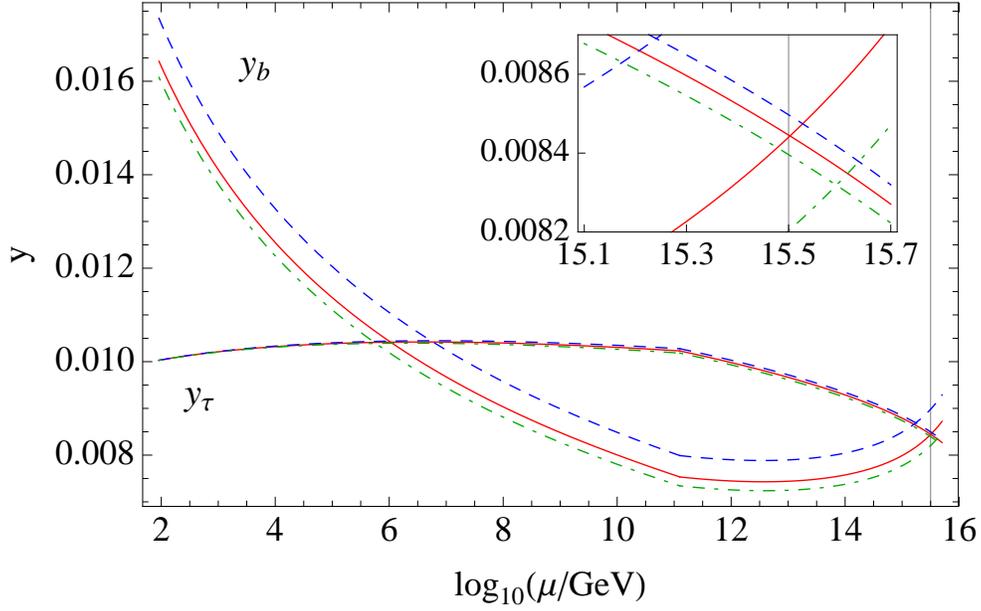}
\caption{Running of $b$ and $\tau$ Yukawa couplings in the case $N_g=1$, $M_N=10^{11.1}$GeV, $M_G=10^{15.5}$GeV. $y_\nu(t_N)=1.58$ is calculated by Eq. (\ref{eyntn}). Dashed and dot-dashed lines show errors given in Ref. \cite{Xing:2011aa}. Vertical lines are drawn at $\mu=M_G$.}
\label{fbt1}
\end{center}
\end{figure}

\begin{figure}[htb]
\begin{center}
\includegraphics[width=13cm]{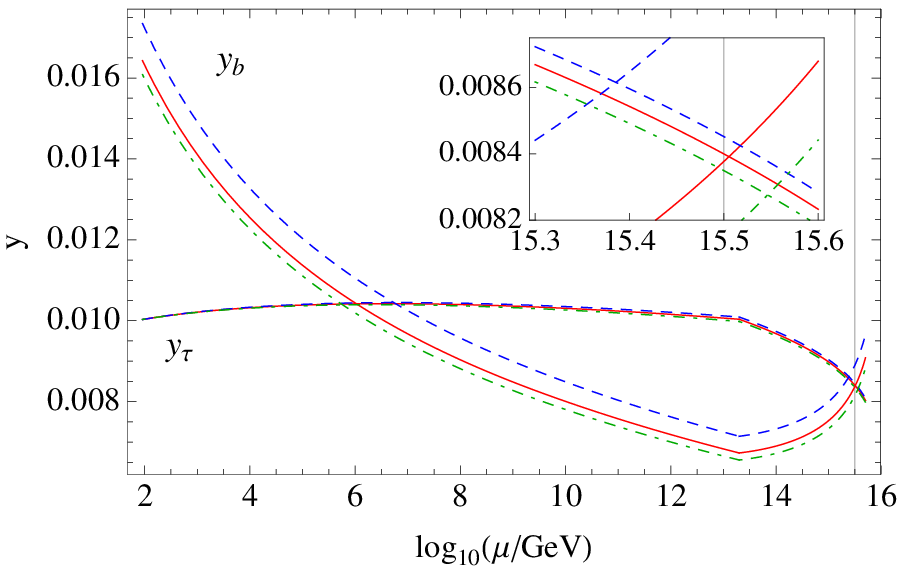}
\caption{Running of $b$ and $\tau$ Yukawa couplings in the case $N_g=2$, $M_N=10^{13.3}$GeV. Other conditions are same as in Fig. \ref{fbt1}.}
\label{fbt2}
\end{center}
\end{figure}

\begin{figure}[htb]
\begin{center}
\includegraphics[width=13cm]{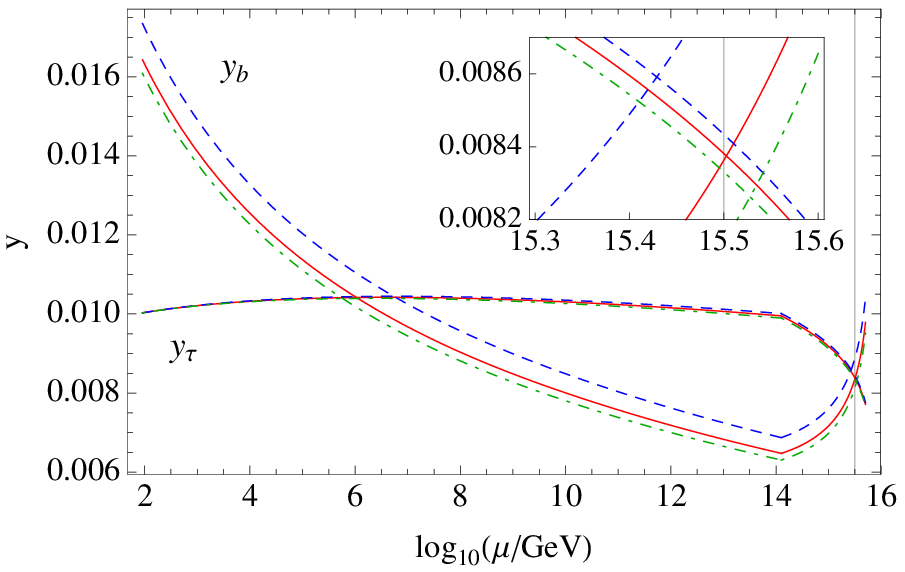}
\caption{Running of $b$ and $\tau$ Yukawa couplings in the case $N_g=3$, $M_N=10^{14.1}$GeV. Other conditions are same as in Fig. \ref{fbt1}.}
\label{fbt3}
\end{center}
\end{figure}

We derive an approximate formula for the initial condition $y_\nu (t_N)$ that satisfies the $b-\tau$ unification condition,
\begin{align}
y_b(t_G)=y_\tau(t_G). \label{ebtau}
 \end{align}
 From Eqs. (\ref{eyb})-(\ref{eyn}), we obtain
\begin{align}
16\pi^2\frac{d}{dt}\ln\frac{y_\nu y_b^2}{y_\tau^2}=\frac{11}{2}N_g y_\nu^2+\frac{71}{20}g_1^2-\frac{9}{4}g_2^2-16g_3^2. \label{enubtau}
 \end{align}
We neglected $y_b$ and $y_\tau$ on the right hand side of Eq. (\ref{enubtau}), since they are much smaller than other couplings. Here we consider the case $M_N>m_{O_F}$. In this region, $y_\nu^2\gg g_1^2,g_2^2,y_t^2$, so Eq. (\ref{eyn}) can be approximately solved,
\begin{align}
y_\nu(t)=\left(y_\nu(t_N)^{-2}-\frac{5N_g}{16\pi^2}(t-t_N)\right)^{-1/2}.
 \end{align}
We found that the difference between this equation and the numerical solution is $\lesssim 1\%$. By using this equation and integrating Eq. (\ref{enubtau}), the GUT relation (\ref{ebtau}) gives the initial condition of $y_\nu$:
\begin{align} 
y_\nu(t_N)=
\left[
\frac{16\pi^2}{5N_g(t_G-t_N)}\left(
1-\left(
\frac{y_b(t_N)}{y_\tau(t_N)}\right)^\frac{10}{3}
\left(\frac{\alpha_1(t_N)}{\alpha_1(t_G)}\right)^{-\frac{71}{24b_1}}
\left(\frac{\alpha_2(t_N)}{\alpha_2(t_G)}\right)^{\frac{15}{8b_2'}}
\left(\frac{\alpha_3(t_N)}{\alpha_3(t_G)}\right)^{\frac{40}{3b_3'}}\right)\right]^{1/2}. \label{eyntn}
\end{align}
This is the key formula in this work. $y_\nu(t_N)$ is plotted in Fig. \ref{fynmn}. By solving RGEs (\ref{eytop}) to (\ref{eyn}) and using the initial condition (\ref{eyntn}), we found that $y_b$ and $ y_\tau$ are unified within the experimental errors. Examples are shown in Figs. \ref{fbt1}-\ref{fbt3}.

$y_\nu(t)$ increases monotonically, so we take a perturbativity bound to $y_\nu(t_G)<\sqrt{4\pi}\;(\sim 3.54)$. It gives an upper bound on the initial condition, 
\begin{align} 
y_{\nu}(t_N)<\left(\frac{1}{4\pi}+\frac{5N_g}{16\pi^2}(t_G-t_N)\right)^{-1/2}.
\end{align}
 This upper bound determines the maximal values of $M_N$ (see Fig. \ref{fynmn}). Those values and $y_\nu(t_N)$ are listed in Table~\ref{tMNmax}.
 
\begin{table}[htb]
\caption{Maximal masses of the right-handed neutrinos and neutrino Yukawa couplings that realize $b-\tau$ unification. ``Central'' is the case of all experimental parameters taken to the best-fit values. ``Best'' indicates the case of maximal $y_b(M_Z)$, minimal $y_t(M_Z)$ and $\alpha_3(M_Z)$. ``Worst'' is the opposite case to ``Best''.}
\begin{center}
\begin{tabular}{ccccccc}\hline\hline
 &\multicolumn{3} {c}{$\log_{10}(M_N/{\rm GeV})$}&\multicolumn{3} {c}{$y_\nu(t_N)$}\\ 
 $N_g$ & Worst & Central & Best& Worst & Central & Best\\ \hline
\multicolumn{4}{l}{($M_G=10^{15.5}$GeV)}&&&\\
1 & 10.7 & 11.1 & 12.1&1.53&1.58&1.75\\
2 & 13.1& 13.3&13.8&1.53&1.58&1.75\\
3 & 13.9&14.1&14.4&1.53&1.62&1.77\\ 
\multicolumn{4}{l}{($M_G=10^{16.0}$GeV)}&&&\\
1 & 10.4 & 10.9 & 12.0&1.43&1.49&1.64\\
2 & 13.2& 13.5&14.0 &1.43&1.50&1.64\\
3 & 14.1&14.3&14.7&1.42&1.49&1.66\\ 
\hline\hline
\end{tabular}
\end{center}
\label{tMNmax}
\end{table}
 
Right-handed neutrinos induce light neutrino mass matrix. The (3, 3) component of the matrix can be expressed as
 \begin{align}
m_{\nu 33}= -\frac{y_\nu(t_N)^2v^2}{M_N}\sum_{I=1}^{N_g} e^{2i\theta_I} +m_{\nu 33}',
  \end{align}
where we have defined $Y_{\nu I3}\equiv y_\nu e^{i\theta_I}$ and $m_\nu'$ is possibly an induced mass by other fields such as $T_F$. The most natural case seems to be the $N_g=3$ case because $M_N$ can be  close to the seesaw scale (see Table \ref{tMNmax}). Even if $M_N\ll 10^{15}$GeV, we can obtain $|m_{\nu 33}|=O$(0.01-0.1)eV by tuning $\theta_I$ or $m_\nu'$.  

 We do not check the stability of the electroweak vacuum here. If $y_\nu$ is $O(1)$, it may make Higgs quartic coupling $\lambda$ negative for $t>t_N$. On the other hand, there are many scalar fields in the SU(5) model because we need adjoint scalar $ 24_H$ to break SU(5) down to $G_{\rm SM}$. Those scalar fields contribute  positively to the running of $\lambda$ \cite{Forshaw:2003kh,Boughezal:2010ry,Haba:2014sia}, so the stability depends on their couplings. $T_F$ and $T_H$ make $\alpha_2$ larger and also contribute to $\lambda$ in the positive direction \cite{Gogoladze:2008ak} (this effect may be small). The vacuum stability is not trivial in our model.
 
{\it Summary.} We have studied the effect of the neutrino Yukawa couplings $y_\nu$ on the running of $y_b$ and $y_\tau$. We have found the $y_\nu(t_N)$ and $M_N$ that can satisfy the GUT relation $y_b(t_G)=y_\tau(t_G)$. If three right-handed neutrinos exist, $b-\tau$ unification is possible with their masses close to the seesaw scale.
 
\section*{Acknowledgments}
 The author is grateful to Takehiko Asaka for helpful discussions.


\clearpage
{}

\end{document}